\documentclass{iucr}              
\papertype{FA}               

\begin{document}                  
\title{New phasing method based on the principle of minimum charge}
\shorttitle{New phasing method}
\cauthor[a]{Pavel}{Kalugin}{kalugin@lps.u-psud.fr}{}
\aff[a]{Laboratoire de Physique des Solides, Universit\'e Paris-Sud, 91405 Orsay \country{France}}
\shortauthor{Kalugin}
\maketitle                        

\begin{synopsis}
Description of a new phasing method and demonstration of its use for the
crystal and quasicrystal structure determination.
\end{synopsis}

\begin{abstract}
A new method of the phase determination in X-ray crystallography is proposed.
The method is based on the so-called "minimum charge" principle, recently
suggested by Elser. The electron density function $\rho$ is sought in the form
$\rho({\bf x})=|\psi({\bf x})|^2$, where $\psi$ is an 
$n\mbox{-component}$ real function. The  norm $\int|\psi({\bf x})|^2 d{\bf x}$ 
is minimized under the
constraint imposed by the measured data on the amplitudes of Fourier harmonics
of $\rho$. Compared to the straightforward implementation of the "minimum
charge" scheme, the method attenuates the Gibbs phenomenon and is also
capable of extrapolation of the diffraction data beyond the set of measured
amplitudes. The method is applicable to quasicrystals under the condition 
that the number of components $n$ of the function $\psi$ is bigger than the 
dimensionality of the ``atomic surface''. It is successfully 
tested 
on synthetic data for Fibonacci chain and the octagonal tiling. In the latter 
case the reconstructed density map shows the shape of the atomic surface,
despite relatively low data resolution. 
\end{abstract}

\section{Introduction}

The vast majority of the existing direct methods of X-ray structure
determination approach the phase problem as a problem of constrained 
minimization.
The quantity to minimize plays the role of the likelihood functional,
optimization of which is subject to constraints imposed by the known
structure factor amplitudes.
The choice of this quantity is a matter of tradeoff between three requirements.
First, this functional should approximatively represent the common 
notion about the likelihood of a given density function. Second, it should be
effectively computable, and finally it should allow for efficient minimization.
Usually none of these requirements is fully satisfied. 
Consider for instance the traditional direct methods based on the 
conditional probability distribution of the structure invariants
\cite{harker,karle}.
The derivation of the expressions for the conditional
probability takes as a starting point the assumption that unconditional
probability of the density distribution (its {\em Bayesian prior}) is a translationally invariant measure on the ensemble of 
$N$ $\delta\mbox{-like}$ atoms in a unit cell 
\cite{cochran}.
Clearly this approach misses the physical constraint on a
minimal distance between the atoms.
Although this prior measure allows one to express the 
conditional probability of the
phase invariants in a closed form \cite{cochran_sym,hauptman,cochran}, 
the resulting formulae are effectively
computable only for invariants of small order. As a result, instead of the 
true conditional probability function the practical algorithms use some 
sort of {\em ad hoc} approximation, usually based on combination of the 
triplets and
quartets (see e.g. ``Shake 'n Bake'' algorithm \cite{snb}). 
However, even these simplified functionals are not easy to optimize.
The existing phase refinement methods are based on iterative procedures and can
easily get stuck in a local minimum of a functional.

Recently, \citeasnoun{elser_orig}
suggested the so-called {\em principle of minimum charge}.
According to this principle, the correct set of phases should minimize the
average charge density 
(the unknown Fourier component with ${\bf k}=0$, which has to be 
added to the density to make
it non-negative). The rationale behind this principle is that the
functions satisfying it tend to have shallow highly degenerated minima and
spiky maxima, which is what one would expect of an atomistic density
function.
Compared to the solid statistical foundations of the conventional direct
methods, the principle of minimum charge may look arbitrary. However,
this comparison is not fair, because on the way from the first principles to
the practical implementation of the conventional methods 
many {\em ad hoc} assumptions are made. 
In contrast, the principle of minimum charge is almost readily 
applicable in the 
phase refinement algorithms. Additional advantage of this principle is that it
could be used without modification for the determination of structure of
quasicrystals and non-commensurate crystals. This is especially important,
because the conventional methods fail when applied to these structures. Indeed,
a na{\"\i}ve attempt to approximate quasicrystals by crystals with very big 
number $N$ 
of atoms in the unit cell 
gives rise to the divergence of the normalized structure factors 
$E_{\bf H} =O(N^{1/2})$ and of the triplet amplitudes $A_{\bf HK}=O(N)$.
This immediately leads to meaningless results (e.g. that phases 
of all triplets $\Phi_{\bf HKL}=0$).    

A straightforward implementation of the principle of minimum charge (e.g. that 
proposed by \citeasnoun{elser_orig}) implies solving a minimax problem.
Indeed, the set of phases $\Phi_{\bf K}$ satisfies the principle if the deepest
minimum of the ``density with zero average''
\begin{equation} 
\label{naive_fourier}
\rho'({\bf r})=2\sum_{\bf K}|F_{\bf K}|\cos({\bf K}\cdot{\bf r}-\Phi_{\bf K})
\end{equation}
over ${\bf r}$ has the maximal possible value over all possible sets 
$\Phi_{\bf K}$. Robust algorithms for finding a global 
saddle point are much more difficult to design 
than those finding a global minimum or maximum. The reason is that the
methods which are usually applied to prevent the algorithm from getting stuck
with a local minimum (e.g. simulated annealing or multiple runs) do not work in
the case of local saddle points. In this article we propose a version of the
principle of minimum charge which can be formulated as a problem of global
minimization, and not as a minimax problem, allowing one to circumvent the
shortcomings of a straightforward approach.

The na\"{\i}ve implementation of the principle of minimum charge suffers from
another drawback, which is due to the so-called Gibbs phenomenon. In order to
understand this problem, consider the case when the true values of all phases 
$\Phi_{\bf K}$ are known. Suppose also, that the correction for the atomic form
factors is included in the values of the structure factors $F_{\bf K}$ in 
(\ref{naive_fourier}), or in other words that the atoms are point-like. 
Due to the Gibbs phenomenon, any finite sum of the form (\ref{naive_fourier})
presents negative ``bumps'' around the positions of atoms. The depth of these
artifacts may be as big as 20\% of the peak hight. This picture is clearly 
different from multiply degenerate shallow minima that one would expect of a
function satisfying the principle of minimum charge. In other words, 
the correct set
of phases gives a density function (\ref{naive_fourier}) which is suboptimal
from the point of view of this principle. Any further
optimization of it may only introduce errors in the values of phases. The
proposed method significantly attenuates the
importance of the Gibbs phenomenon, although does not remove it completely.

\section{Method}
\subsection{Representation of the density function}

\begin{figure}
\label{degenerate_minima}
\caption{Phase refinement based on the principle of minimum charge leads to
the reconstructed density maps with highly degenerate shallow minima.}
\scalebox{0.37}{\includegraphics{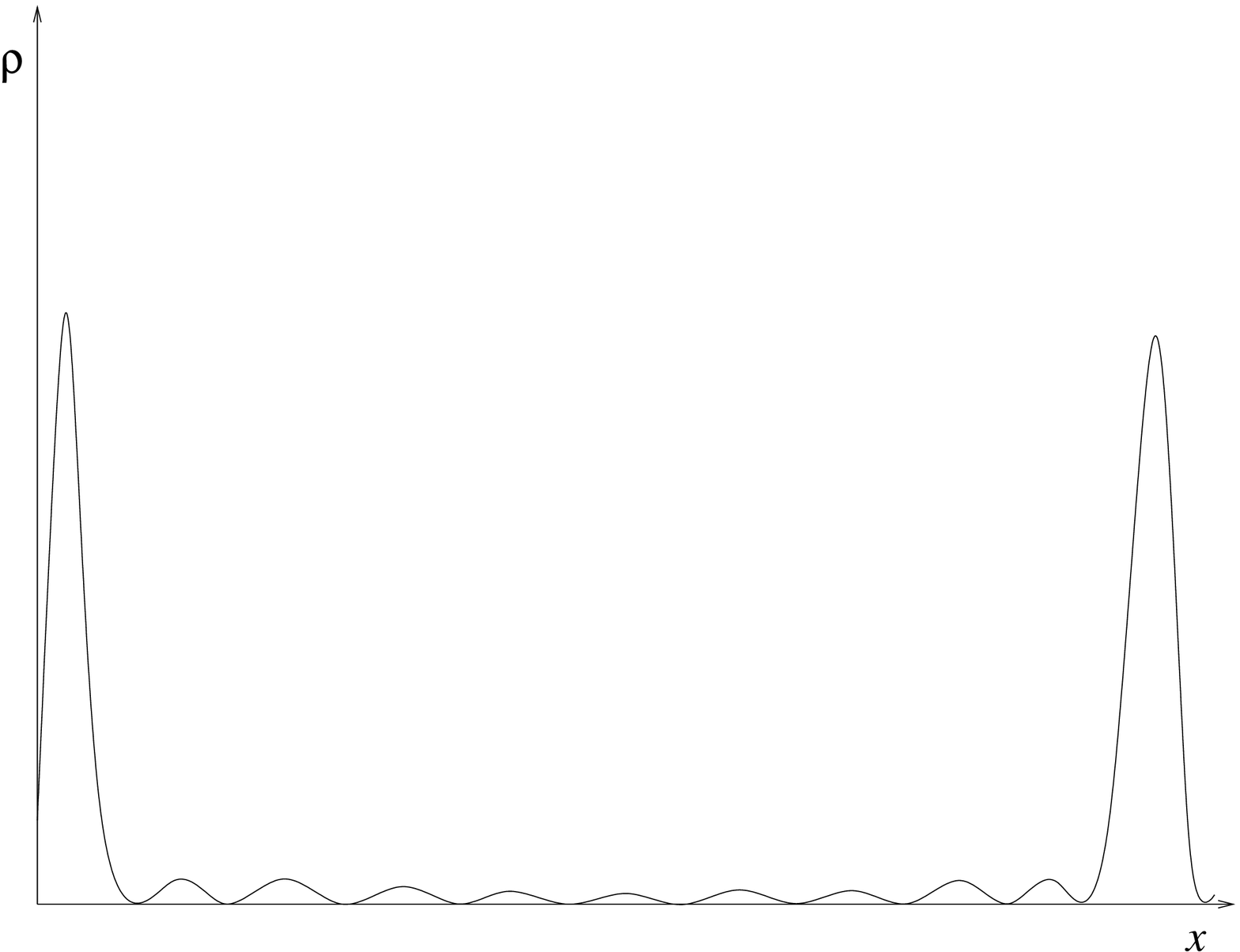}}
\end{figure}

The cornerstone of the new method is the representation of the atomic density
function. The traditional way of reconstructing this function consists of
using finite Fourier sums (\ref{naive_fourier}) with possible application of
weights aimed to attenuate Gibbs phenomenon. A different approach is used here.
The density function is approximated by {\em a square} of a finite Fourier sum.
More precisely, the density function $\rho({\bf x})$ is modeled as
\begin{equation}
\label{psisq}
\rho({\bf x})=|\psi(\bf x)|^2,
\end{equation} 
where $\psi(\bf x)$ is a multi-component real function. Each component
$\psi_\alpha$ of $\psi$ has a finite Fourier spectrum:
\begin{equation}
\label{psi}
\psi_\alpha({\bf x}) = \sum_{{\bf K} \in \Lambda} 
\tilde\psi_{\alpha, {\bf K}}
e^{i {\bf K} \cdot {\bf x}},
\end{equation} 
where $\Lambda$ is a finite subset of reciprocal lattice vectors. The complex
coefficients $\tilde\psi_{\alpha, {\bf K}}$ satisfy the condition
\begin{equation}
\label{conj}
\tilde\psi_{\alpha, -{\bf K}}=\tilde\psi_{\alpha, {\bf K}}^*
\end{equation}
Before going any further, let us discuss the rationale behind this 
representation. Recall, that the density function satisfying the principle of
minimum charge should possess the following properties:
\begin{enumerate}
\item \label{fidelity} The amplitudes of its Fourier components for the 
set of measured Bragg reflections should correspond to the measured data
$F_{\bf K}$.
\item \label{atomicity} The global minimum of the function should be 
highly degenerate (see Fig. \ref{degenerate_minima}).
\end{enumerate}
The traditional way to represent the density function by a finite Fourier sum
automatically guarantees that the condition \ref{fidelity} is satisfied.
However, the condition \ref{atomicity} cannot be expressed in a closed form
as an explicit constraint on the phases $\Phi_{\bf K}$. On the other hand,
degenerate global minimum occurs naturally in the functions of the type 
(\ref{psisq}) when $\psi({\bf x})$ has multiple zeros. In the same time,
the condition \ref{fidelity} can be explicitly
imposed on the function (\ref{psisq}). This constraint can de formulated
as a system of equations of 4-th order on the variables 
$\tilde\psi_{\alpha, -{\bf K}}$ (see Appendix A).

Mention should be made of the criterion for choice of 
the support $\Lambda$ for the Fourier
spectrum of the function $\psi$ in the equation (\ref{psi}). 
Clearly, if the
set $\Lambda$ is too small, the condition \ref{fidelity} may be impossible to
satisfy. This happens e.g. if there are wave vectors ${\bf L}$ 
belonging to set $\Xi$ of measured
data, which cannot be represented as ${\bf L}={\bf H}-{\bf K}$, where 
${\bf H}, {\bf K} \in \Lambda$. On the other hand, choosing the set 
$\Lambda$ equal to the set $\Xi$ guarantees that any constraint on the
amplitudes of Fourier components of density $\tilde\rho_{\bf K}$
\begin{equation}
\label{rho_fourier}
\tilde\rho_{\bf K}=\int \rho({\bf x}) e^{-i {\bf K}\cdot{\bf x}} d {\bf x}
\end{equation}
with ${\bf K}\in \Xi$ can be satisfied in the representation (\ref{psisq}).
Indeed, assuming the following constraint on the amplitudes of $\tilde\rho_{\bf K}$:
$$
|\tilde\rho_{\bf K}|=a_{\bf K},
$$
setting
$$
\tilde\psi_{\alpha, {\bf K}}=
\left\{
\begin{array}{ll}
\epsilon^{-1} & \qquad\mbox{if }{\bf K}=0,\quad \alpha = 0\\
\frac{\epsilon}{2}a_{\bf K}&\qquad\mbox{if }{\bf K}\neq 0,\quad \alpha = 0\\
0&\qquad\mbox{if }\alpha \neq 0
\end{array}
\right.
$$
gives the correct Fourier components of $\rho$ in the limit 
$\epsilon \rightarrow +0$:
$$
|\tilde\rho_{\bf K}|=a_{\bf K}+O(\epsilon).
$$
Note, that in this case the support of the spectrum of 
$\rho$ in this case is roughly 
twice as large in the reciprocal space as the set $\Lambda$. In other words, the
representation (\ref{psisq}) allows for extrapolation of the structure factors.
One can further extrapolate the experimental data by choosing an even
larger set $\Lambda$. The limit of possible improvement of the resolution in
this way is set by the occurrence of spurious peaks in the density map in the
case of over-extrapolation due to the increased sensitivity of the result to the
errors in the measured data.

\subsection{Optimization criterion and constraints}

The representation (\ref{psisq}) of the density is a non-negative function. 
Hence the average value of $\rho$ can be used as a figure of merit for
the optimization, in other words, the principle of minimum charge regains 
its original meaning. It should be emphasized, that instead of the problem of
finding a global saddle point (global maximum of global minima) we deal here
with an easier problem of constrained minimization. The role of
parameters is played by the Fourier components $\tilde\psi_{\alpha, {\bf K}}$ of
the function $\psi$. The average density is expressed through 
$\tilde\psi_{\alpha, {\bf K}}$ as
\begin{equation}
\label{averdens}
\langle \rho \rangle = 
\sum_{{\bf K} \in \Lambda}
\sum_\alpha |\tilde\psi_{\alpha, {\bf K}}|^2
\end{equation}
The parameters $\tilde\psi_{\alpha, {\bf K}}$ can be considered as a real vector
in the $Mn\mbox{-dimensional}$ space, where $M$ is the number of points in the
set $\Lambda$ and $n$ is the number of components of $\psi$. The principle of
minimal charge thus boils down to the problem of constrained minimization of 
the norm of this vector.

\begin{figure}
\label{gibbs_artefact}
\caption{Artifacts occurring due to Gibbs phenomenon.
The original structure is a one-dimensional crystal with one atom
per unit cell. 
The number of independent structure factors is 10.}
\scalebox{0.5}{\includegraphics{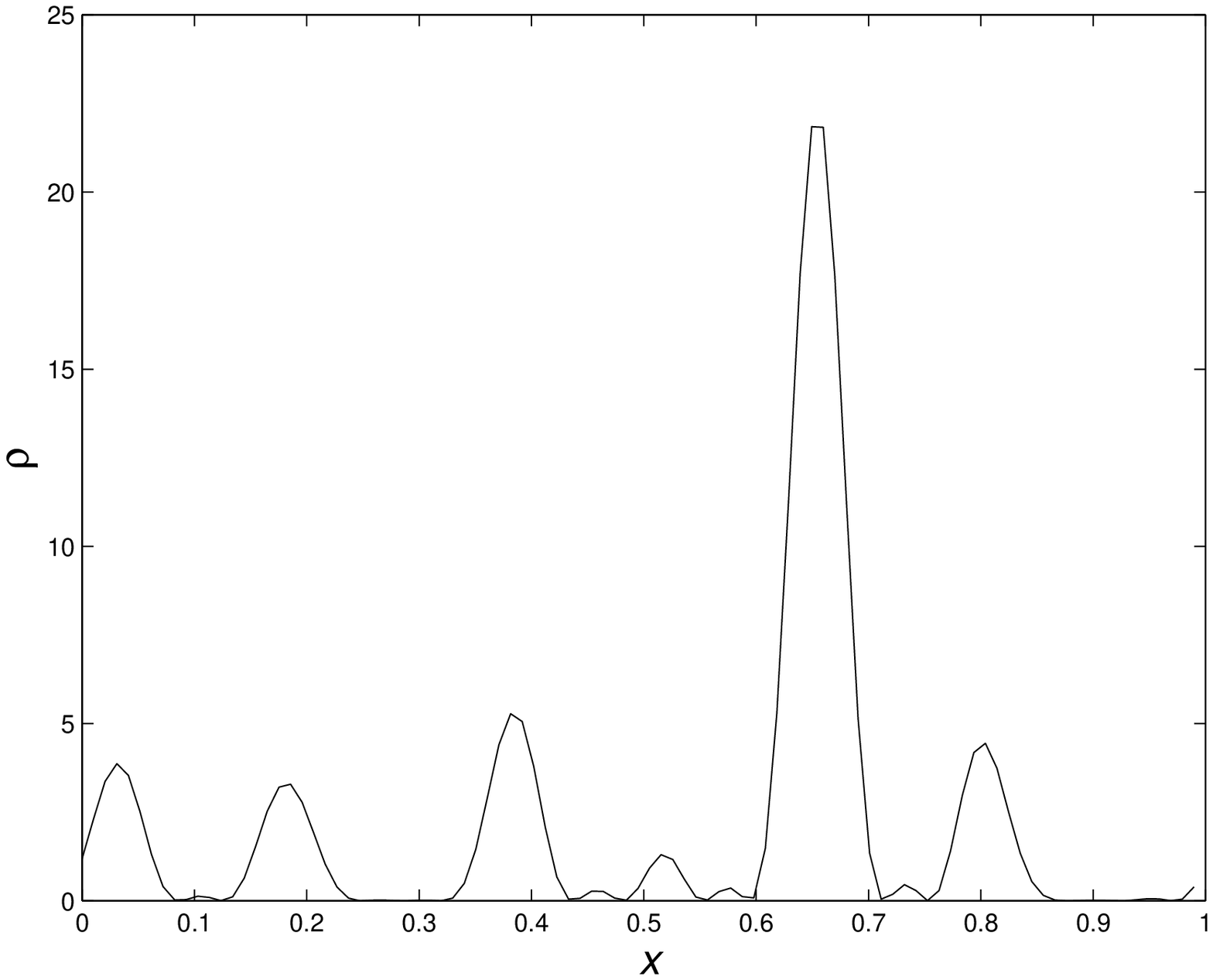}}
\end{figure}

Consider now the constraints imposed by the measured data on the values
$\tilde\psi_{\alpha, {\bf K}}$. Let us assume that the amplitudes of the 
structure factors 
$|F_{\bf K}|$ are measured for the wave vectors ${\bf K}$ belonging to a 
subset $\Xi$ of the  reciprocal lattice. Then the seemingly obvious constraint
on the density function $\rho({\bf x})$ from (\ref{psisq}) consists of 
the requirement that
$$
|\tilde \rho_{\bf K}|=|F_{\bf K}|
\qquad\mbox{ for any }{\bf K}\in\Xi.
$$
This condition expressed in terms of the Fourier
components of $\psi$ from (\ref{psi}) takes the form:
\begin{equation}
\label{noweights}
\left|
\sum_\alpha
\sum_{{\bf H}, {\bf K}-{\bf H} \in \Lambda} 
\tilde\psi_{\alpha, {\bf H}}\tilde\psi_{\alpha, {\bf K}-{\bf H}}
\right|=|F_{\bf K}|
\qquad\mbox{ for any }{\bf K}\in\Xi.
\end{equation}
This constraint, however, does not take into account the effect of the Gibbs
phenomenon. The importance of this effect is clear from the Figure 
\ref{gibbs_artefact}. 

This plot shows the function $\rho({\bf x})$ obtained by
the minimization of the average density (\ref{averdens}) under the constraints
(\ref{noweights}). The structure factors used here correspond to a one-dimension
crystal with one atom per unit cell, that is $|F_{\bf K}|=1$ for all 
${\bf K}\in\Xi$.
Clearly the phasing on the Figure \ref{gibbs_artefact} is incorrect.
The reason is that the correct set of phases would correspond to a function with
deep negative bumps around the ``atom'' because of truncation of data 
in the reciprocal
space. The optimization routine attempts to
flatten these bumps out and align all minima on the same level, giving rise to a
wrong solution.

The standard approach to reduction of the Gibbs phenomenon consists of using the
windowing in the Fourier domain. As applied to the formula (\ref{noweights}) the
windowing consists of multiplying the amplitudes of the structure factors by
weights:
\begin{equation}
\label{withweights}
\left|
\sum_\alpha
\sum_{{\bf H}, {\bf K}-{\bf H} \in \Lambda} 
\tilde\psi_{\alpha, {\bf H}}\tilde\psi_{\alpha, {\bf K}-{\bf H}}
\right|=w_{\bf K}|F_{\bf K}|.
\end{equation}
The choice of the coefficients $w_{\bf K}$ is a matter of tradeoff between 
softening the truncation of data on the reciprocal lattice and preserving the
spatial resolution of the density map. 
In the field of digital signal processing there
exist many windowing formulae designed to reduce the magnitude of Gibbs 
phenomenon, e.g. Welch or Hann windows \cite{DSP,Bloomfield}. Nonetheless, 
instead of using
these formulae, we shall derive the expressions for $w_{\bf K}$ which are
optimized from the point of view of the discussed method. 
Ideally, the weights $w_{\bf K}$ should guarantee that the global minimum of 
the average density (\ref{averdens}) with the constraints 
(\ref{withweights}) corresponds to the density map $\rho$ with the correct
phases. The weights
should be a function of the sets $\Xi$ and $\Lambda$ 
only, and should not depend on the values of the structure factors $F_{\bf K}$.
Unfortunately, such ideal weights do not exist. This is clear at least
from the fact that the perfect phasing is impossible when the set of measured
data is too small as compared to the number of atoms in the unit cell.
Nevertheless, as long as the resolution of the measured data is sufficiently
high, there exists a nearly optimal set of weights. 
More specifically, the weights 
$w_{\bf K}$ chosen in such a ways as to guarantee the correct phasing for
the structure with one atom per unit cell should produce reasonable results for
other structures as well, provided that the peaks in the density 
map do not overlap.

Let us start the construction of the optimal weights with the case of a
one-dimensional crystal with one atom in a unit cell. 
First of all, note that the constraints in the formula (\ref{noweights}) with
$|F_{\bf K}|=1$ are equivalent to those of the equation (\ref{withweights})
with $|F_{\bf K}|=1/w_{\bf K}$. In other words, one can interpret the 
density map on the Fig. \ref{gibbs_artefact} as a result of correct phasing of
an a priori unknown structure factor amplitudes $|F_{\bf K}|=1/w_{\bf K}$. 
Clearly, the number of peaks in this density function depends on the number of
constraints in (\ref{noweights}). 
One would naturally expect that the minimization of the global charge
with only one constraint leads to some very simple structure. 
Indeed, as shown below, if the set $\Xi$ in (\ref{noweights}) contains only one
wavevector ${\bf K}_0$,
which is equal to the elementary period of the reciprocal lattice,
the minimization of the average density gives rise to a
structure with one atom in a unit cell.  Let 
$\tilde\psi_{\alpha, {\bf K}}=\tilde\eta_{\alpha, {\bf K}}$ be
the solution of (\ref{noweights}) with the single constraint
$|F_{{\bf K}_0}|=1$. One can use these
values to construct the following set of weights:
\begin{equation}
\label{weights_fit}
w_{\bf K}=
\frac{
\left|
\sum_\alpha
\sum_{{\bf H}, {\bf K}-{\bf H} \in \Lambda} 
\tilde\eta_{\alpha, {\bf H}}\tilde\eta_{\alpha, {\bf K}-{\bf H}}
\right|}
{\sum_{{\bf K} \in \Lambda}
\sum_\alpha |\tilde\eta_{\alpha, {\bf K}}|^2}.
\end{equation}
This set of weights guarantees that the density map obtained with a single
constraint will satisfy the principle of minimum charge for any number of
constraints.
Indeed, $\tilde\psi_{\alpha, {\bf K}}=
w_{{\bf K}_0}^{1/2}\tilde\eta_{\alpha, {\bf K}}$ 
obeys the equations
(\ref{withweights}) with the weights (\ref{weights_fit}) and $|F_{\bf K}|=1$ for
all ${\bf K}\in \Xi$. In other words, the weights $w_{\bf K}$ are optimal for
the set $\Lambda$.

\begin{figure}
\label{single_atom}
\caption{Reconstructed density map for the same test case as that of the Fig.
\ref{gibbs_artefact}, but with constraints (\ref{withweights}). The values of
weights are given by the formula (\ref{one-d-weights}).}
\scalebox{0.5}{\includegraphics{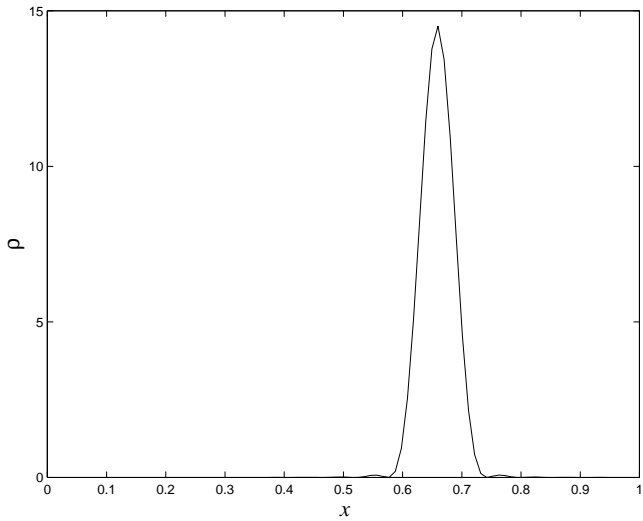}}
\end{figure}

Let us actually compute the optimal weights for the case when the set 
$\Lambda$ includes the nodes $-N, -N+1, \dots N$ of the one-dimensional 
reciprocal lattice.
We suppose that $\psi({\bf x})$ has only one component and use
simplified notations by writing $\tilde\psi_k$ or $F_k$
instead of $\tilde\psi_{\alpha, {\bf K}}$ or $F_{\bf K}$
(here $k$ is the number of the node). 
In the case when only the amplitude of the structure factor
$F_1$ is known, there is only one constraint on the values $\{\tilde\psi_i\}$:
\begin{equation}
\label{one-d-constraint}
\left|
\sum_{k=-N+1}^N \tilde\psi_k \tilde\psi^*_{k-1}
\right|
= w_1 |F_1|
\end{equation} 
By introducing Lagrange multiplier the constrained minimization 
of the average density can be replaced by finding an
unconstrained extremum of the following quantity:
\begin{equation}
\label{lagrange}
\sum_{k=-N}^N \tilde\psi_k \tilde\psi^*_k
+\lambda \left|
\sum_{k=-N+1}^N \tilde\psi_k \tilde\psi^*_{k-1}
\right|
\end{equation}
Note, that the second sum in the formula (\ref{lagrange}) can be always  made
real and positive by an appropriate shift $\delta x$ 
of the coordinate system and
multiplication of $\tilde\psi_k$ by $e^{ik\delta x}$.
Then if expression (\ref{lagrange}) has an extremum at a given
$\{\tilde\psi_i\}$, the same is true for the expression
\begin{equation}
\label{lagrange2}
\sum_{k=-N}^N \tilde\psi_k \tilde\psi^*_{k}
+\frac{\lambda}{2} \sum_{k=-N+1}^N (\tilde\psi_k \tilde\psi^*_{k-1} + 
\tilde\psi_k^* \tilde\psi_{k-1})
\end{equation} 
Differentiation with respect to $\tilde\psi$ gives the equation 
for the point of extremum:
\begin{equation}
\label{laplace}
\tilde\psi_k+\frac{\lambda}{2}(\tilde\psi_{k-1}+\tilde\psi_{k+1})=0
\end{equation}
(here we assume $\tilde\psi_{-N-1}=\tilde\psi_{N+1}=0$). 
In other words, the solution of the
principle of minimum charge with the constraint (\ref{one-d-constraint}) should
be an eigenstate of a lattice Laplace operator with zero boundary
conditions beyond the set $\Lambda$. Straightforward calculations show that the
following function
\begin{equation}
\label{psimin}
\psi_k=C \cos\left(\frac{\pi k}{2(N+1)}\right) 
\end{equation}
gives the smallest value of the average density. The density is a sum of peaks
centered at the lattice nodes
\begin{equation}
\label{sum_of_peaks}
\rho(x)=\sum_{n \in Z} f[(2N+2)(x-n)],
\end{equation}
where the shape of an individual peak is given by the following formula:
\begin{equation}
\label{peak}
f[u]=C \left(
\frac{\cos(\pi u)}{1-4 u^2}
\right)^2.
\end{equation}
As one might expect, in the limit $N\rightarrow\infty$ the density map tends to
a sum of $\delta\mbox{-functions}$ corresponding to a crystal with one atom per
unit cell.
The formula (\ref{weights_fit}) gives the values for the optimal
weights:
\begin{equation}
\label{one-d-weights}
w_k=\frac{
\sin\left((2N+3-|k|)\theta\right) + (2N+3-|k|)\sin(\theta) \cos(k\theta)
}
{
\sin\left((2N+3)\theta\right) + (2N+3)\sin(\theta) \cos(k\theta)
},
\end{equation}
where $\theta=\pi/2(N+1)$. The results of testing of these weights with the
structure factors corresponding to one atom in the unit cell are shown on Figure
\ref{single_atom}.

Consider now the ways to generalize the above construction of the 
optimal weights to
more than one dimension. In this case the minimal number of constraints needed
to define a structure is equal to the dimensionality $d$ of the
crystal. Solving the problem of minimum charge with the known structure factors
for ${\bf K}={\bf K}_1 \dots {\bf K}_d$
yields to the formula similar to
(\ref{laplace}):
\begin{eqnarray}
\label{laplace_multi_d}
\tilde\psi_{\alpha, {\bf K}}+
\sum_i
\frac{\lambda_i}{2}
(\tilde\psi_{\alpha, {\bf K}-{\bf K}_i}+
\tilde\psi_{\alpha, {\bf K}+{\bf K}_i})=0\\
\tilde\psi_{\alpha, {\bf K}}=0\qquad\mbox{ for }{\bf K}\notin\Lambda.\nonumber
\end{eqnarray}
The weights can be obtained from the solution $\tilde\psi_{\alpha, {\bf K}}$ of
this equation using the formula (\ref{weights_fit}).

Contrary to the one-dimensional case, equation 
(\ref{laplace_multi_d}) does not
yield to a unique set of
optimal weights. First of all, the choice of the wave vectors ${\bf K}_i$ is
ambiguous. By analogy with the 
one-dimensional crystal, it seems natural to require
that the vectors ${\bf K}_i$ form a basis of the reciprocal lattice. However,
there are many ways to choose a basis of a lattice in more than one dimension.
The other problem is related with the choice of Lagrange multipliers
$\lambda_i$. In the formula (\ref{laplace}) the solution $\psi_k$ is always one
of the eigenstates of Laplace operator and the role of the parameter $\lambda$
is restricted to mere selection of the eigenstate. In contrast,  
$\tilde\psi_{\alpha, {\bf K}}$ in (\ref{laplace_multi_d}) may vary continuously 
with $\lambda_i$. In order to resolve the ambiguity one needs to recall that
the density corresponding to the solution of the finite difference equation 
(\ref{laplace_multi_d}) describes the structure with one atom at the
origin of the unit cell. In other words, the function $\psi(\bf x)$ should have
a sharp peak at origin, that is its Fourier components 
$\tilde\psi_{\alpha, {\bf K}}$ should vary slowly across the domain $\Lambda$. 
The requirement that the equation (\ref{laplace_multi_d}) should
admit of such slowly
varying solution determines the choice of parameters ${\bf K}_i$ and 
$\lambda_i$.


\subsection{Quasicrystals and incommensurate structures}

Consider now the application of the discussed method to the deterministic
quasicrystals and incommensurate structures in general. These structures can be
conveniently described using the so-called superspace or 
``cut-and-project'' method
\cite{superspace,cut}. According to this approach, the density function of quasicrystals
can be obtained as a cut through a periodic function in a space of higher
dimensionality. The Fourier spectrum of the structure is obtained
as a projection of the spectrum of the periodic function on the cut direction,
and thus consists in a discrete sum of $\delta\mbox{-functions}$. If the
direction of the cut is incommensurate with the periodicity, 
each $\delta\mbox{-peak}$ in the Fourier spectrum of the quasicrystal
corresponds to a node of the reciprocal lattice of the periodic function.
By this means the phase problem for quasicrystals can be reformulated in a
conventional way as a problem of phasing for a periodic function in a space of
higher dimensionality.

For the case of point-like atoms the density function of a quasicrystal 
in the real space is a
discrete sum of  $\delta\mbox{-functions}$. The corresponding periodic density
function consists of $\delta\mbox{-like}$ distributions on sub-manifolds,
commonly referred to as ``atomic surfaces'' \cite{at_surf,bak}. This brings up 
again
the question of optimality of the weight factors $w_{\bf K}$ from 
(\ref{withweights}). Indeed, the weights obtained following the method 
described in the previous section are appropriate for distributions of narrow
non-overlapping peaks. Atomic surfaces clearly do not fall into this
category. As a result one could expect that using the weight factors optimized
for point-like atoms might give rise to incorrect results in the case of atomic
surfaces. Nevertheless, lacking any better alternative, we applied these weights
for quasicrystals as well. Despite concerns, the results of the numerical tests 
described below show no significant distortion of the density map.

Another problem emerges 
when the atomic surfaces are discontinuous, which is the case for all known
deterministic structure models of real materials. The point is that due to the
finite resolution in the reciprocal space, the boundary of the atomic surface is
inevitably smoothed out. As a result, when the physical space cuts through an
atomic surface near its boundary, the amplitude of the corresponding peaks in
the density map is reduced. Similarly, when the cut misses the atomic surface by
a short distance, a ``phantom'' peak appears in the density map. In reality both
of the above phenomena occur simultaneously, because of the so-called
``closedness'' property of the atomic surfaces \cite{closesurf}. This results in
occurrence of double peaks of reduced intensity, which are often separated by a
distance much smaller than the typical interatomic spacing. A careful
analysis shows that such double peaks in places group together to form more
complex clusters \cite{jumpcluster}. 
It should be emphasized that all these artifacts are 
not specific to the phasing method and will exist in any reconstructed 
quasicrystalline density map. They should not be confused with true partially
occupied sites which occur in real quasicrystals because of the phason disorder
\cite{lyon,ara,diffuse}.

\subsection{Choice of the number of components of $\psi$}

\begin{figure}
\label{trapped}
\caption{An unremovable hole in the atomic surface on the reconstructed density
map. Such holes occur when a line of zeros of one-component function $\psi$
crosses the atomic surface.}
\scalebox{0.4}{\includegraphics{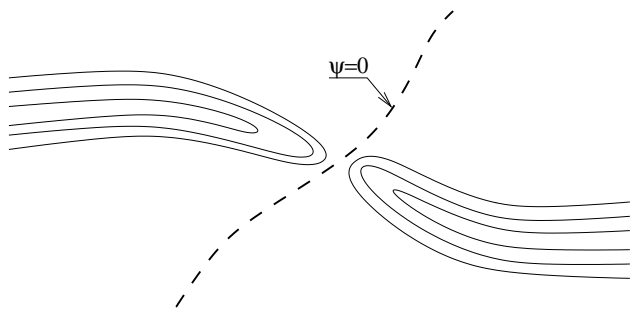}}
\end{figure}

To this point the number of components of the function $\psi({\bf x})$ has been
of no importance for us. However, because of the iterative nature of the
optimization algorithm, this parameter plays an important role in the case of
quasicrystalline structures. The problems which arise in the case when the
number of components of $\psi$ is too small are clear from Fig.~\ref{trapped}.
This figure depicts a one-dimensional atomic surface 
in the two-dimensional space crossed 
by a line of zeros of one-component function $\psi$. The resulting hole 
can only be removed by pushing it towards the boundary of the atomic surface. 
It may occur, however, that pushing the hole inwards decreases the average
density. In this case, the algorithm will converge to a local minimum.

The above problem could be avoided if the function $\psi$ had two components. In
this case, the zeros of $\psi$ are points. Would such point fall onto the atomic
surface, it could be easily pushed away by a small perturbation of $\psi$. The
same applies to the case of more than two dimensions. In this case the number of
components of $\psi$ should be bigger, than the dimensionality of the atomic
surface. 

\section{Results}

The results presented in this section were obtained numerically using the  
algorithm of constrained minimization described in the Appendix A. 
The algorithm was tested on the sets of synthetic data including one-dimensional
crystal and quasicrystal, as well as a two-dimensional quasicrystalline
structure. No rotational
symmetry was assumed in any of the cases. When the structure actually possessed
additional symmetry, it was recovered as a result of optimization. 

Because of its iterative nature, the algorithm of Appendix~A does not guarantee
convergence to the global minimum of the average density. It is worth
noting, however, that for all tested structures
the correct phasing corresponds to the deepest of the found minima.

\subsection{One-dimensional crystals}

The test structure includes five point-like atoms. Their `charges'
and fractional coordinates $x$ are given in the Table~\ref{five_atoms}. 
As the structure does not possess central symmetry, the reconstructed density
can correspond to any of the enantiomorphs with equal probability.
\begin{table}
\label{five_atoms}
\caption{One-dimensional test structure}
\begin{tabular}{@{}l@{\hspace{10em}}r@{}}
`charge' &  $x$ \\
2.0 & 0.0 \\
1.5 & 0.6 \\
1.2 & 0.25 \\
1.0 & 0.43 \\
1.0 & 0.8
\end{tabular}
\end{table}

\begin{figure}
\label{12amplitudes}
\caption{The restored density map corresponding to the structure from 
Table~\ref{five_atoms}. The input data includes first 12 structure factor
amplitudes. The function $\psi$ has two components and is defined with the same
resolution.}
\scalebox{0.5}{\includegraphics{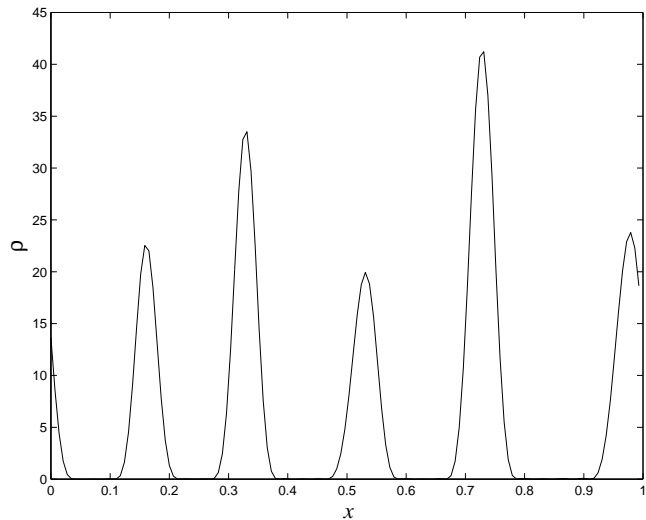}}
\end{figure}

The density map shown on the Fig. \ref{12amplitudes} has been obtained with the 
first 12 structure factor amplitudes. The function $\psi$ had 2 components. The
support of the Fourier spectrum of each component $\Lambda$ from the formula
(\ref{psi}) included the wavevectors from the interval $[-12, 12]$. The
positions of five peaks on the Fig.~\ref{12amplitudes} correspond to the
coordinates of atoms in the Table~\ref{five_atoms} up to a global translation.
The amplitudes of peaks are also qualitatively recovered. One can remark,
however, that the two smallest peaks on the Fig.~\ref{12amplitudes} are 
noticeably different, although they correspond to identical
 atoms. The value of the
average density is equal to $6.636$, which is about $1\%$ smaller than the total charge
$6.7$ of the Table~\ref{five_atoms}. 
The iterative optimization converged to the global minimum solution 
in 14 of 20 trials when starting with random
normally distributed $\tilde \psi_{\alpha, {\bf K}}$ was $70\%$, giving the 
$70\%$ success rate.

\begin{figure}
\label{extrapolation}
\caption{Density map for the test structure of the Table~\ref{five_atoms}
extrapolated from the low resolution data. }
\scalebox{0.5}{\includegraphics{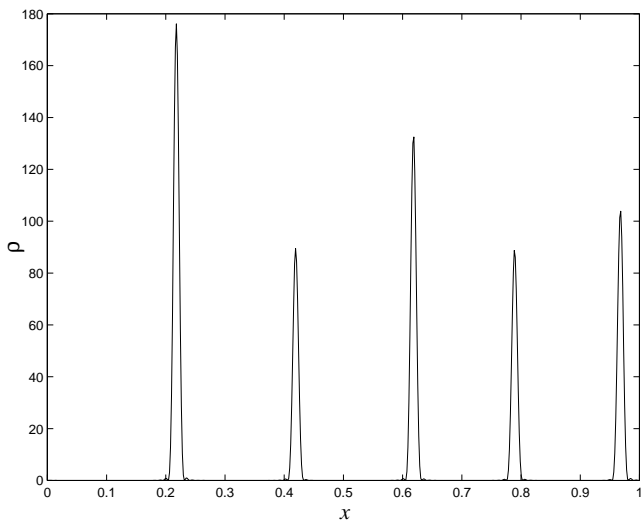}}
\end{figure}

The test structure from the Table~\ref{five_atoms} was also used to benchmark
the extrapolating capabilities of the algorithm. The input data was restricted
to the first 9 structure factor amplitudes. Note, that the 
structure recovery with
any smaller data set would be impossible because a one-dimensional
structure of 5 atoms is described by 9 real parameters not including a global
translation. By contrast, the set $\Lambda$ from (\ref{psi}) included all
wavevectors from the interval $[-50, 50]$. By this means, the algorithm
extrapolated the ``experimental'' scattering data to a region of the reciprocal
space roughly 11 times larger that that where they were initially defined. The
reconstructed density map is shown on the Fig.~\ref{extrapolation}. The
positions of atoms are perfectly accurate up to a global translation and
inversion. Note also, that despite a smaller resolution of the input data 
(which is only about $0.65$ of the 
smallest interatomic distance), the amplitudes of
the peaks are restored with higher accuracy than on the Fig.\ref{12amplitudes}.
The value of the average density is equal to $6.690$, which is also much closer to
the total charge $6.7$ of the Table~\ref{five_atoms}. This can be explained by 
overlapping of wider peaks on the Fig.\ref{12amplitudes}, which makes the
correct solution suboptimal. As a result, the algorithm produces slightly
distorted solutions, which have smaller values of the average density. Narrower
peaks on the  Fig.\ref{extrapolation} make this phenomenon much less visible.

One could anticipate a lower success rate in the latter example because 
of the bigger
number of parameters in $\tilde \psi_{\alpha, {\bf K}}$ to fit. 
Contrary to the expectations, 
convergence to the global minimum was obtained in 19 trials of 20.

\subsection{One-dimensional Fibonacci chain}

\begin{figure}
\label{cut-and-project}
\caption{The ``cut-and-project'' method of construction of the Fibonacci chain. 
The one-dimensional physical space cuts the two-dimensional lattice of 
periodically arranged segments (``atomic surfaces''). The atoms, 
shown as filled circles, are located at
intersections of ``atomic surfaces'' with the physical space.}
\scalebox{0.35}{\includegraphics{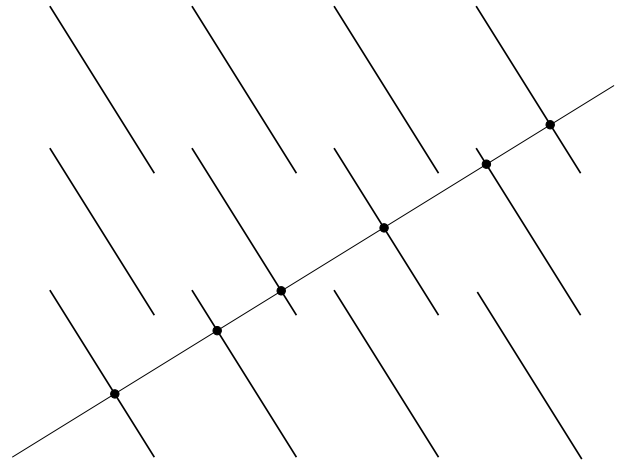}}
\end{figure}

Fibonacci chain is probably the best known example of a one-dimensional
quasicrystal. The chain consists of identical atoms arranged on a straight line.
The distance between neighbouring atoms may take two values, usually referred 
to as ``short'' and ``long'' segments of the chain. The ratio of lengths of
short and long segments equals $\tau=(\sqrt{5}-1)/2$. The short and long
segments alternate following a quasiperiodic Fibonacci sequence, which is
usually defined recursively \cite{fibrec}. The same structure can be obtained
following the conventional ``cut-and-project'' method 
(see Fig.~\ref{cut-and-project}). 

\begin{figure}
\label{fibonacci}
\caption{Reconstruction of the atomic surfaces of the one-dimensional Fibonacci
chain. The input data includes 40 independent 
reflections with $h^2+k^2\le25$. 
The function $\psi$ has two components.
The shown
area of the density map includes 4 unit cells. The axes represent fractional
coordinates.}
\scalebox{0.6}{\includegraphics{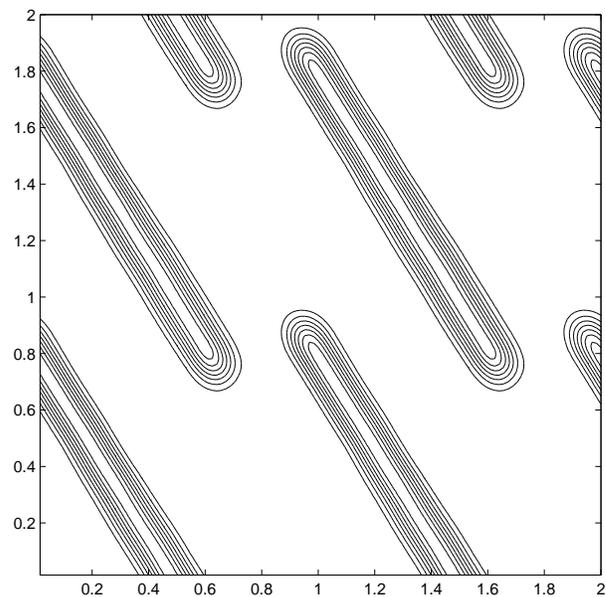}}
\end{figure}

The algorithm has been tested on a set of 40 independent reflections with the 
Miller indices $h^2+k^2 \le 25$. Although the structure on the 
Fig.~\ref{cut-and-project} possesses central symmetry, 
this symmetry was not
imposed as an additional constraint on $\psi$. The reconstructed density map is
shown on the Fig.~\ref{fibonacci}. The optimization has yielded the correct
solution with $100\%$ success rate for 50 trials with random initial conditions.

\subsection{Two-dimensional octagonal Ammann-Beenker tiling}

Ammann-Beenker octagonal tiling \cite{ammann,beenker} refers to quasiperiodic
covering of a plane by squares and $45^\circ$ rhombi. 
Structure models associated with this tiling can be constructed by decorating
each tile with atoms. The simplest decoration consists of placing atoms at
vertices of squares and rhombi. The resulting structure can be obtained by
``cut-and-project'' technique from a periodic density function in
four-dimensional space. The corresponding atomic surfaces are two-dimensional
perfect octagons, one per unit cell. 

The input data included the reflections with $h^2+k^2+l^2+m^2 \le 5$ (where 
$h$, $k$, $l$ and $m$ stand 
for four-dimensional Miller indices). There are 68 pairs of opposite nodes of
the reciprocal lattice satisfying the above inequality. The
$8mm$ symmetry of the diffraction pattern makes only 14 of them
independent. However, as mentioned above, the point symmetry was not taken into
account. As a result all 68 pairs of reflections were
considered as independent. The function $\psi$ had three components. Note, that
as no central symmetry is imposed, the coefficients 
$\tilde\psi_{\alpha, {\bf K}}$ from the formula (\ref{psi}) are complex numbers.

The optimal set of coefficients $\tilde\psi_{\alpha, {\bf K}}$ returned by 
the algorithm should be converted to the density map in the physical space.
As the direction of the physical space is incommensurate with the
four-dimensional periodic lattice, this conversion implies Fourier transform of
unevenly spaced data. This precludes using standard FFT algorithms, which
significantly slows down displaying the results.
A workaround for this problem consists of tilting the physical space slightly to
make it commensurate with the lattice. This is equivalent to replacing the
quasicrystal by a close approximant \cite{approx}. 

\begin{figure}
\label{ammann_restored}
\caption{Density map corresponding to the vertices of an approximant to 
Ammann-Beenker tiling, 
reconstructed from the reflections with $h^2+k^2+l^2+m^2 \le 5$. }
\scalebox{0.65}{\includegraphics{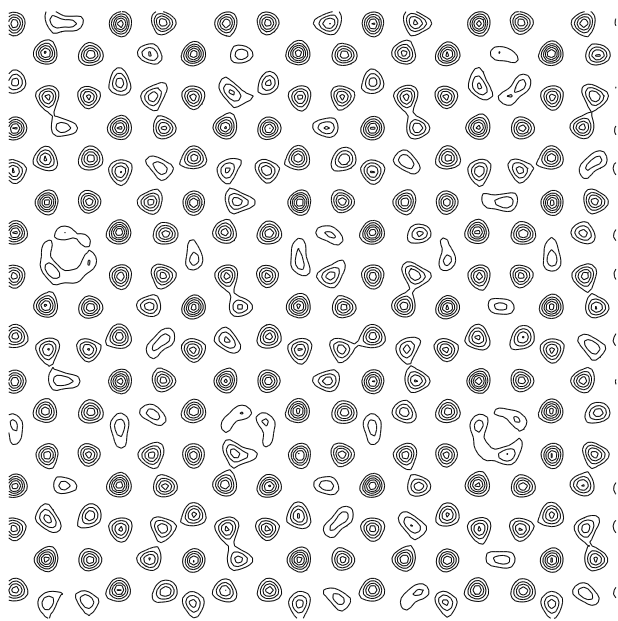}}
\end{figure}
 
\begin{figure}
\label{ammann_original}
\caption{The vertices of the approximant to Amman-Beenker tiling with the
physical space spanned by the vectors (\ref{approx}).}
\scalebox{0.66}{\includegraphics{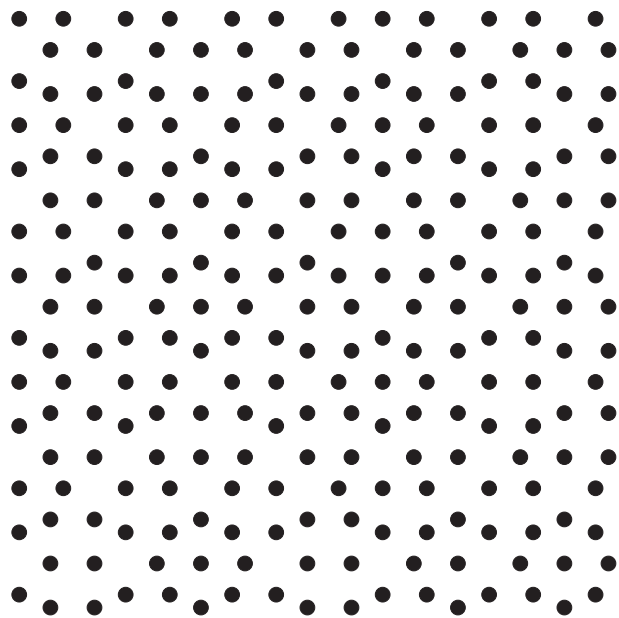}}
\end{figure}

Figure~\ref{ammann_restored} depicts the reconstructed density map for an
approximant to Ammann-Beenker tiling. The approximant is obtained by replacing 
the vectors  ${\bf u}_1=(\sqrt{2}/2, 1/2, 0, -1/2)$ and  
${\bf u}_2=(0, 1/2, \sqrt{2}/2, 1/2)$ spanning the physical space by
the rational vectors:
\begin{equation}
\label{approx}
\begin{array}{l}
{\bf u}_1'=(5/7, 1/2, 0, -1/2)\\  
{\bf u}_2'=(0, 1/2, 5/7, 1/2). 
\end{array}
\end{equation}
Note that the peaks 
on the density map substantially overlap because of low resolution of the 
input data. This is also confirmed by the fact that the algorithm yields the
average density as low as 0.718 of its correct value, suggesting
significant over-optimization. Nevertheless, most of the atomic positions are
correctly resolved, as can be seen from comparison with the vertices of 
the ideal approximant tiling shown on Fig.~\ref{ammann_original}. The double
peaks which are expected due to the smoothened edges of the atomic
surface (see the explanation above) are not fully resolved. They manifest
themselves as elongated features on the density map. The optimization algorithm 
has converged to the correct solution with $100\%$ success rate of 10 trials 
with random initial conditions.

\begin{figure}
\label{octagon}
\caption{The cut through the reconstructed density map of Ammann-Beenker tiling
in the direction parallel to the atomic surface. The scale is arbitrary.}
\scalebox{0.6}{\includegraphics{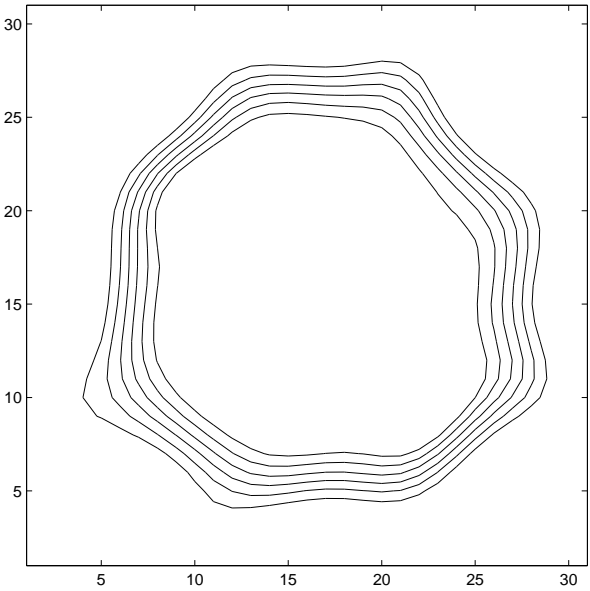}}
\end{figure}

One could also consider cutting the four-dimensional density function along
different directions. For instance, a cut in the direction of the atomic surface
would reveal its shape. In practical conditions this shape will be
distorted because the cut may not pass exactly through the center of the
four-dimensional peak corresponding to the atomic surface. 
Such cut is shown on Fig.~\ref{octagon}. Despite low resolution of the input
data one can clearly see a faceted octagonal shape. 

\section{Summary and discussion}

We have described a new method of structure determination based on the
minimum charge principle. The method possesses extrapolating capabilities, and
can restore atomic positions from low resolution data (about $65\%$ of the
interatomic distance). It is also applicable to quasicrystals and incommensurate
structures. As the current implementation of the algorithm was intended as a
demonstration of principle, no special efforts were taken to make it
more efficient. Making the new method practical for crystals with
non-trivial point symmetry and for real quasicrystal requires further research.

The failure to take advantage of the symmetry of the structure is a major
drawback of the current implementation. The problems related to the symmetry are
the most visible in the case of central symmetric structures.
It is well known, that in the case of central symmetry, 
the structure factors $F_{\bf K}$ are real numbers 
(the same applies to some structure factors when
the point symmetry group contains other elements of order 2). As a result, each
of the constraints (\ref{withweights}) on the absolute values of the Fourier
components of $\rho({\bf x})$ defines two disconnected manifolds 
in the $nM\mbox{-dimensional}$ space of coefficients 
$\tilde\psi_{\alpha, {\bf K}}$. The total number of disconnected manifolds
defined by all constraints is equal to $2^m$, where $m$ is the number of real
structure factors. 
In its current implementation, the optimization algorithm sticks to one of
these pieces in the early stages
and and then continues looking for the point of minimal average 
density on this piece only. 
In this way, the global minimum will most likely be missed. 

The other problem is related to the fact that the 
symmetry of the function $\psi({\bf x})$ does not necessarily 
coincide with that of the density function $\rho({\bf x})$. In the general case,
rotations and translations in the real space are accompanied by orthogonal
transformations in the $n\mbox{-dimensional}$ space of components of $\psi$. In
other words, the symmetry group of $\psi({\bf x})$ is an extension of the
symmetry group of $\rho({\bf x})$ by a subgroup of $O(n)$. Generally speaking, 
the search for the solution corresponding to the global minimum of the average 
density should be performed over all such extensions.

\appendix
\section{Numerical algorithm}

This section describes in details the algorithm used to minimize 
$|\psi|^2$ with the constraints (\ref{withweights}).
The algorithm converges quadratically 
in the vicinity of a local minimum with the
computational complexity of one iteration $O((Mn)^3)$
and the storage requirement of the order $O((Mn)^2)$. Here 
$M$ stands for the number of points in the set $\Lambda$ in (\ref{psi})
and $n$ is the number of components of $\psi({\bf x})$.
These properties are suitable for refining an already found approximate
minimum. In other words, this algorithm should 
be used as a second stage in a two
stage optimization scheme. Nonetheless, the results of testing described above
show that this algorithm can work also in a single stage, starting with random
values of parameters.

The minimal charge problem can be stated in the following abbreviated form:
\begin{eqnarray}
\mbox{minimize } |\psi|^2\nonumber\\
\label{constr}
\mbox{subject to } h_i(\psi)=c_i.
\end{eqnarray}
Here $\psi$ is a vector the real $Mn\mbox{-dimensional space}$. 
The minimization
of its norm is subject to $N$ nonlinear equality constraints (\ref{constr}) 
representing the
conditions (\ref{withweights}). Note, that by squaring the both sides of 
(\ref{withweights}) $h_i$ can be chosen in the form of uniform polynomials of
4-th degree in $\psi$.  By this means all derivatives of 
$h_i$ are readily available in an analytic form.
This makes appropriate application of the sequential quadratic programming (SQP)
techniques of optimization \cite{sqp1,sqp2}. We have used the following scheme
for an elementary SQP iteration:
\begin{enumerate}
\item\label{start}{
Compute the SVD decomposition of the Jacobian $J=\nabla_\psi h$ of constraints: 
$J=USV$, where $UU^T=\hat 1$, $VV^T=\hat 1$ and $S$ is an $N\times nM$ diagonal
matrix.
Denote the first $N$ rows of $V$ by $V_{\mathrm{range}}$ and the resting rows by
$V_{\mathrm{null}}$. These matrices give the projectors correspondingly 
onto the range and null spaces of the constraints.
}
\item{
Compute Newton step in the range space:
$$
\delta\psi_{\mathrm{range}} =  V_{\mathrm{range}}^T S^{-1} U^T \cdot (c-h),
$$
where $S^{-1}$ stands for a square $N\times N$ diagonal matrix of inverse
singular values.
}
\item{
Compute the row vector of approximate Lagrange multipliers:
$$
\lambda=-2 \psi^T V_{\mathrm{range}}^T S^{-1} U^T
$$
}
\item\label{hessian}{
Compute the Hessian of the Lagrangian function $L = \psi^T \psi + \lambda h$:
$$
H=\frac{\partial^2 L}{\partial\psi \partial\psi}.
$$
}
\item\label{projection}{ 
Compute the projection of $H$ onto the null space of the constraints
$H_{\mathrm{null}}=V_{\mathrm{null}} H V_{\mathrm{null}}^T$.
}
\item\label{quadratic}{
Find the search step $\delta\psi_{\mathrm{null}}=V_{\mathrm{null}}^T x$ in 
the null space. Ideally, this step should minimize the quadratic 
approximation to the variation of the Lagrangian:
$$
\delta L \sim 2\psi^T V_{\mathrm{null}}^T x + \frac{1}{2}x^T H_{\mathrm{null}} x.
$$ 
Note that the quadratic form $H_{\mathrm{null}}$ is not 
necessarily positive definite.
}
\item\label{update}{ Update $\psi$:
$$
\psi := \psi + \delta\psi_{\mathrm{null}} + \delta\psi_{\mathrm{range}}
$$
}
\item{Repeat the steps \ref{start}-\ref{update} until convergence.}
\end{enumerate}
The storage requirements of the algorithm are dominated by the necessity to
store a dense $nM \times nM$ Hessian matrix $H$ and its projection for 
the steps \ref{hessian}-\ref{quadratic}. The speed bottleneck are the steps
\ref{projection} and \ref{quadratic}, both requiring $O((nM)^3)$
multiplications. The SVD decomposition at the step \ref{start} and the 
computation of Hessian at the step \ref{hessian} take $O(N^2(nM))$ and
$O(N(nM)^2)$ multiplications correspondingly. Taking into account that 
$M$ scales as $N$, the contribution of these steps to the computation time
may be non-negligible for small $n$.

The only non-trivial point in the algorithm consists of handling possibly 
non-positive definite Hessian matrix at the step \ref{quadratic}. The Hessian
often has non-positive eigenvalues when the trial point is far from the local
minimum. Note also, that for the considered problem the local minimum is
always degenerate because of translational invariance in the real space and
rotational invariance in the space of the components of the field 
$\psi({\bf x})$. As a result, the Hessian at the local minimum always has 
at least $d+n-1$ zero eigenvalues. The standard approach to this problem
consists of modifying the Hessian matrix to make it positive definite. We use
the fact that diagonalization of a symmetric matrix is numerically stable to
represent $H_{\mathrm{null}}$ in the form
$$
H_{\mathrm{null}}=\sum_i \mu_i \nu_i \nu_i^T,
$$
where $\nu_i$ form an orthogonal basis in the null space and $\mu_i$ are the
eigenvalues of $H_{\mathrm{null}}$. The modified Hessian
matrix is then given by the formula
$$
\tilde H_{\mathrm{null}}=\sum_i \min\{|\mu_i|, \epsilon\} \nu_i \nu_i^T,
$$
As the matrix $H_{\mathrm{null}}$ is dimensionless, the regularization parameter
$\epsilon$ can be set to some constant numeric value, which should be much
bigger than the machine epsilon but still small enough to keep the convergence
quadratic in the vicinity of the minimum. The matrix $\tilde H_{\mathrm{null}}$
is then used to compute the full Newton update step in the null space:
$$
x=-2 \tilde H_{\mathrm{null}}^{-1} V_{\mathrm{null}} \psi.
$$

\ack{The author is grateful to E.~Tatarinova for fruitful discussions.}

\referencelist[acta2001]
\end{multicols} 
\end{document}